\documentclass[11pt,a4paper]{emulateapj}
\def\be{\begin{equation}}
\def\ee{\end{equation}}
\def\gal{\textsc{Galform}}
\def\galsp{\textsc{Galform} }

\def\OII{[\ion{O}{2}]}
\def\HA{H$\alpha$ }
\usepackage{graphicx}
\usepackage{natbib}
\usepackage{epsfig}
\usepackage{amsmath}
\begin{document}
\title{The Role of Ram Pressure Stripping in the Quenching of Cluster Star Formation}
%

\author{Laura G.\ Book and Andrew J.\ Benson}
\affiliation{\itshape Mail Code 350-17, California Institute of Technology, Pasadena, CA 91125}

%
%
\begin{abstract}

Recent observations of galaxy clusters have shown that environmental effects apparently associated with the cluster begin to lower the star formation rates of galaxies at distances as great as three times the cluster virial radius. These observations may indicate preprocessing of cluster galaxies in groups or in the cluster core for galaxies on highly elliptical orbits, but may also imply that the environmental effects due to the cluster are directly affecting galaxies on their first infall. To explore these issues, we investigate different models of ram pressure stripping as it acts on satellite galaxies in clusters, and compare to observations of the radial star formation gradient in clusters. We calculate the location of the accretion shock around model clusters, and use this as the radius of onset of ram pressure stripping in the \galsp semi-analytic model of galaxy formation. Comparison of the results of our model, and previously considered, simpler ram pressure models, with recent observations indicates that current data is unable to strongly discriminate between models of ram pressure stripping due to the complex interplay of preprocessing effects at work. However, future observations of a larger sample of clusters will likely be able to place stronger constraints on the process of ram pressure stripping and its role in shaping radial trends in and around clusters.
\end{abstract}
\subjectheadings{cosmology: theory, galaxies: clusters: general, galaxies: evolution, galaxies: clusters: intergalactic medium}

\maketitle
\pagestyle{empty}
%
%
\section{Introduction}  \label{sec:intro}

It is widely known that the star formation rate (SFR) of galaxies depends on the density of their environment, and specifically that galaxies within galaxy clusters tend to be forming stars at a lower rate than comparable field galaxies. Since star formation is fueled by gas, this leads to the conclusion that galaxies in clusters tend to have less gas from which to form stars than their counterparts in less dense environments \citep{BO78}. 

Several mechanisms have been proposed to explain the observed trend of star formation with environment. It is well known that spiral galaxies tend to be bluer, and therefore forming stars at a higher rate, than elliptical and S0 galaxies,  the so-called morphology-density relation \citep{Dressler80}, and that spiral galaxies tend to be rarer in the centers of galaxy clusters than early type galaxies \citep{Whitmore93,Barkhouse09}. It is clear, therefore, that morphological transformation, driven by interactions such as mergers or multiple weak gravitational encounters with other satellites \citep[galaxy harassment, e.g.][]{Moore96}, likely plays a role in driving the observed quenching of star formation in clusters. However, this cannot completely explain the trend as it has been observed that even among galaxies with the same bulge to disk ratio, field galaxies are forming stars at higher rates than cluster galaxies \citep{Balogh99}. 

Gas may be removed from galaxies passing through the dense intergalactic medium of a cluster by means of ram pressure stripping (RPS). As originally envisaged, in this mechanism the interstellar medium (ISM) of a satellite galaxy is lost due to interaction with the dense gas of the host halo through which it is moving \citep{Gunn72, Abadi99, Quillis00, Kenney09}. Alternatively, a satellite moving through less dense material may lose just a portion of its diffuse atmosphere of hot halo gas \citep[colloquially known as starvation or strangulation, e.g.][]{Bekki02, McCarthy08}. Finally, it is inevitable in a hierarchical Universe that many cluster galaxies were previously members of lower mass groups of galaxies, and environmental effects such as strangulation in these lower density environments may have begun the quenching of the star formation of these galaxies before they were accreted onto the cluster \citep{Zabludoff98, McGee09, Kawata08, Fujita04}.

It has traditionally been assumed that these environmental effects only occurred once satellites had fallen through the virial radius of a host halo. However, recent observational studies of satellite galaxy SFR versus cluser-centric radius have found that the SFR remains depressed relative to the field out to $2-3$ times the virial radius of clusters \citep{Balogh00, Verdugo08, Braglia09}. Three explanations have been suggested for the radial extent of environmental effects: that group-scale effects in the locality of the cluster are having a large impact on these cluster galaxies \citep{Zabludoff98}, that many satellite galaxies follow highly elliptical orbits which take them close to the cluster center, where they experience strong RPS, and then back out to large radius \citep{Balogh00}, and that the radius at which cluster environmental effects begins to become important is further out than the virial radius. 

In fact, it is already known that using the virial radius as the location at which environmental processes related to the cluster begin is only an approximation. In current structure formation scenarios, dark matter halos are formed through gravitational collapse when an overdense patch of dark matter stops expanding with the universe and collapses. The surrounding gas then falls into this potential well. If the scale of the halo is large enough so that the gravitational dynamical time of the halo is much less than the cooling time of the gas, then the gas will form an accretion shock, where the kinetic energy of the infalling gas is converted into thermal energy, heating the gas to the virial temperature of the halo \citep{White91}. These are the so-called ``virial shocks," as they are expected to form near the virial radius of the halo.

Accretion shocks were predicted in a cosmological scenario by \citet{Bertschinger85}, who found self-similar, spherically symmetric analytic solutions of collisional gas falling into a density perturbation in an Einstein-de Sitter universe.  More recently, models of large-scale, spherical accretion shocks have been developed by \citet{Furlanetto04} and \citet{Barkana04}, and three-dimensional cosmological N-body and hydrodynamical simulations have confirmed the existence of such shocks \citep{Keshet03, Molnar09}. Further, \citet{Molnar09} has predicted that the extent of such shocks should be observable in the Sunyaev-Zel'dovich effect of clusters with next-generation radio telescopes such as ALMA. While virial shocks may be unstable, and so not survive, around low mass halos, they are expected to be an inevitable result of structure formation for halos above a few times $10^{11} M_{\odot}$ \citep{Birnboim03, Keres05}, including groups and clusters of galaxies. When applied in the \galsp semi-analytic structure formation code, the criterion for shock stability of \citet{Birnboim03} gives the similar result that stable shocks form only in halos more massive than around $10^{12} M_{\odot}$ \citep{Benson10}.

To determine the cosmological importance of accretion shocks, we apply this well-developed theory to galaxy formation, and in particular to the environmental effects on galaxies in clusters. Semi-analytic models of structure formation are a very powerful tool to investigate the impact of this additional physics on structure formation, as their relatively light computational requirements allow us to probe the effects of accretion shocks on large samples of galaxy clusters, therefore complementing the hydrodynamical simulations mentioned above. Such semi-analytic models have proven to be very successful probes of galaxy formation; they have well reproduced the luminosity function of galaxies locally \citep{Cole00, Somerville01, Baugh05}, and at high redshift \citep{Kauffmann99}. The addition of astrophysical effects such as supernova feedback, a photoionizing background, and environmental effects have substantially improved the fit at low luminosities \citep{Benson02b, Benson02a, Somerville02}, while the recent addition of feedback and heating effects to counter large-scale cooling flows, such as heat conduction and feedback from active-galactic nuclei, have produced faithful matches to the high-luminosity tail of the luminosity function \citep{Croton06, Bower06, Somerville08}. The colors of satellite galaxies have also recently been matched by the addition of a more detailed model of RPS (\citealp{Font08}; see also \citealp{Lanzoni05, Weinmann09}).

In this work, we consider the physical and observational consequences of the onset of environmental effects at the radius of the accretion shock rather than around the virial radius as has been previously assumed. To study this problem we apply RPS including accretion shocks to the clusters in the Millennium Simulation \citep{Springel05} using the \galsp semi-analytic model of structure formation. In section \ref{sec:Galform} we describe the \galsp model that we use as the basis of our calculations, while in section \ref{sec:Shocks} we briefly describe our calculation of the location of the accretion shock and its implementation in the semi-analytic model, giving a more detailed description of our calculation in the Appendix. In section \ref{sec:res} we present the cluster galaxy properties produced by our model, and compare them with other models of RPS. In section \ref{sec:obs} we compare our model with recent observational results of the SFRs of cluster galaxies, and in section \ref{sec:disc} we discuss the conclusions that we draw from these results for the observability of environmental effects in clusters.

\section{Simulations}  \label{sec:sim}

\subsection{The GALFORM Semi-Analytic Model}  \label{sec:Galform}

We use as the basis of our work the \galsp semi-analytic model of galaxy formation as described by \citet{Bower06}, to which the reader is referred for a detailed description of the model formulation. As described in \citet[see also \citealp{Helly03}]{Bower06}, we apply this model to the detailed merger histories of the entire volume of the Millennium Simulation \citep{Springel05}. Our model uses the same cosmological parameters as that simulation: $\Omega_{\rm m}~=~0.25$, $\Omega_{\rm b}~=~0.045$, $\Lambda~=~0.75$, and $h~=~0.73$ at $z~=~0$. 

As we are working with N-body merger trees, there are situations in which halos decrease in mass with time \citep{Helly03}, as a result of unbound particles incorrectly being tagged as halo members for example. The original implementation of our semi-analytic model was not well equipped to deal with mass loss in halos, so these N-body merger trees were artificially forced to conserve mass. However, in this work we utilize the merger trees without requiring mass conservation, as described by \citet{Stringer10}, as this is a fairer representation of the true behavior of the N-body simulation and is important for this work in which we utilize halo mass growth rates to compute virial shock radii. We find that relaxing the requirement of mass conservation results in an increase in the mean stellar mass content of galaxies of around $0.3$ dex. This change is larger than expected from the work of  \citet{Helly03} due to the sensitive nature of the AGN feedback included in our current model (but which was not present in that of \citealp{Helly03}).

We therefore found it necessary to adjust a single parameter of our model relative to that of \citet{Bower06} to retain a good fit to the observed local $b_J$- and K-band galaxy luminosity functions. Specifically we adjust the parameter $\alpha_{\rm cool}$, described by \citet{Bower06}, which determines the halo mass scale above which AGN feedback becomes effective. We find that increasing $\alpha_{\rm cool}$ from $0.58$ to $0.9$ reduces the lower mass limit for AGN heating to become effective from a few times $10^{11} M_{\odot}$ in the Bower model to a few times $10^{10} M_{\odot}$ in this work, and therefore reduces the mean stellar mass of galaxies. This change tends to cancel out the change of stellar masses caused by not conserving mass in merger trees (as was done in \citealt{Bower06}), and brings our model back into agreement with the local luminosity functions.

The prescription for the treatment of reheated gas in satellite galaxies that we adopt is the same as that of \citet{Font08}, to which the reader is directed for complete description. In brief, gas in a satellite galaxy that is reheated by supernovae or AGN feedback is transferred to the hot halo of the satellite, from which it is transferred to the host halo as the satellite is ram pressure stripped.

In the \citet{Bower06} model, a galaxy that is identified as part of a friends-of-friends group of a more massive halo is considered its satellite. These tend to roughly correspond to satellites within the virial radius of the halo, although the prescription is by no means spherically symmetric. The satellite has all of its hot gas instantaneously stripped away, leaving the cold gas in the ISM of the galaxy but removing the source of gas to be accreted onto the galaxy and form stars. Although this model has been quite successful in reproducing the luminosity function and star formation history of galaxies and their evolution, it fails to reproduce the colors of satellite galaxies, tending to predict them to be redder than is observed.

The challenge to reproduce the correct colors of satellite galaxies was taken up by \citet{Font08}, who implemented a more nuanced approach to the RPS of satellite galaxies based on the hydrodynamic simulations of \citet{McCarthy08}. Keeping the definition of satellite galaxies as those belonging to the friends-of-friends group of a more massive halo, in their model each satellite is assigned an orbit assuming the velocity distributions determined by \citet{Benson05}, and from this they calculate the maximum ram pressure exerted on the satellite by the host halo and galaxy, which occurs at the pericenter of its orbit. The radius at which this maximum ram pressure is equal to the  gravitational restoring force per unit area of the satellite is termed the stripping radius, and all of the hot halo gas beyond this radius is stripped at the moment that the satellite crosses the virial radius of the host halo. This calculation was repeated (possibly resulting in more gas being removed from the satellite) every time its host halo doubled in mass since the previous ram pressure calculation. This model provides a less extreme implementation of RPS, and manages to match the colors of satellite galaxies by allowing them to accrete hot gas and remain blue for a longer time after being accreted.

However, the ram pressure model of \citet{Font08} continues to use the virial radius as the location around which a satellite begins to feel ram pressure from its host galaxy. In fact, ram pressure forces begin to be felt by the infalling galaxy when it passes through the accretion shock, at which the cluster gas temperature, density and pressure discontinuously increase. The accretion shock can be up to twice as far from the host galaxy as the virial radius, potentially significantly altering the effect of a massive host halo on nearby galaxies. Additionally, the model of \citet{Font08} uses randomly assigned orbital parameters for satellites to compute their orbit and, therefore, the ram pressure force that they experience.

\subsection{Implementation of  Accretion Shocks}  \label{sec:Shocks}

In this work we calculate the location of the accretion shock of halos and use this as the radius at which the RPS of satellite galaxies begins, thereby more completely modeling the environmental effects of a host halo on its satellites. We calculate the radius of the accretion shock with a model based on the calculations of \citet[hereafter V03]{Voit03}, with a few assumptions relaxed to obtain a more accurate accretion radius in a wider range of situations. We use the method of \citet{Font08} to implement the RPS of a galaxy once it comes within this radius of the cluster, but use the actual orbit of the satellite (taken directly from the N-body simulation) to compute the ram pressure force experienced at each timestep of our calculation.

The method of accretion shock calculation of V03 is an approximate solution with many simplifying assumptions. For example, it assumes smooth, spherical accretion, an assumption that is known to be quite incorrect in the context of hierarchical galaxy formation. For this reason, there is uncertainty as to the accuracy of the predictions of this model for the hydrostatic structure of cluster gas. For example, cosmological hydrodynamic simulations have shown that the hot gas halo of a cluster can extend well beyond the location of the accretion shock \citep{Frenk99}, pointing to far more complicated physics than is included in the simple V03 model. This discrepancy is likely to be small, as the relative velocity between infalling satellite galaxies and accreting gas, which determines the ram pressure along with the density, will tend to be low outside the shock radius since they are both falling into the host halo and feel little pressure, while within the shock radius the accreted gas becomes nearly stationary creating a large relative velocity with the satellites. However, as it has been shown in numerical simulations of the hierarchical dark matter and gas evolution of galaxy clusters that the profiles of the outer regions of clusters tend to agree with smooth accretion models \citep{Borgani09}, we use this simplified model in this work. 

The V03 calculation of the accretion shock radius also assumes that the accretion shock is always perfectly strong, so that the Mach number of the shock approaches infinity. We relax the strong-shock assumption to allow the accretion shock to have any strength, which we find to be justified as even some massive clusters have accretion shocks with Mach numbers $\mathcal{M} \sim 10$. We also correct for the neglect of an integration constant in the hydrostatic profile derived for the clusters, which we found to be non-negligible.

Briefly, the calculation of the accretion shock proceeds as follows: using the shock jump conditions and the assumption that the total accreted gas mass must be contained within the accretion shock, we obtain a simple hydrostatic model of the cluster gas. Using this model we derive an equation for the accretion radius in terms of the halo mass, mass accretion rate and halo concentration, as well as the Mach number of the accretion shock. By also assuming that the cluster gas has adiabatically contracted from a temperature of $T_{\rm IGM} = 3000 K$ in the intergalactic medium, we simultaneously solve for the Mach number of the accretion shock and its position.

The calculation on which our model is based is described in Appendix A of V03, and a more detailed description of our generalized version of the calculation can be found in the appendix of this paper.

\section{Cluster Galaxy Properties with Different Models of Ram Pressure Stripping}  \label{sec:res}

We first compare the properties of galaxies in our model, in which the stripping of gas begins at the radius of the accretion shock, with the similar model of \citet{Font08} in which these effects begins at the virial radius, and with the model of \citet{Bower06} in which all of a satellite galaxy's hot gas, not just the maximum amount as determined by the parameters of its orbit within the host halo, is stripped away at the virial radius. We will call these models the Shocks, Font, and Bower models respectively.

\begin{figure}[t!]
\includegraphics[width=8.5cm]{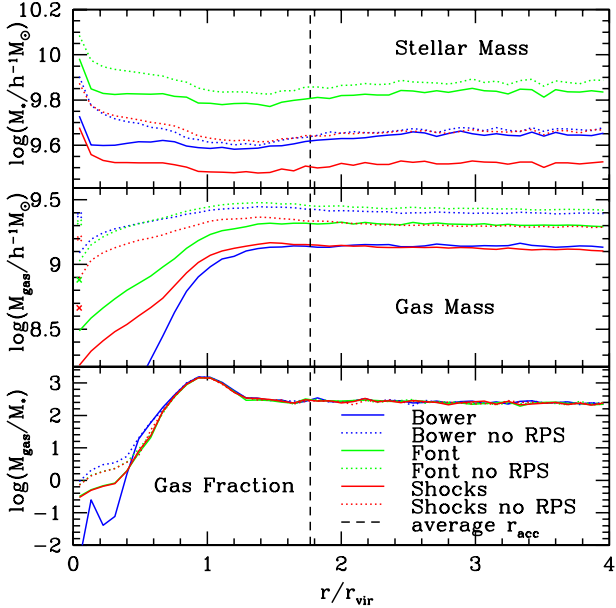}
\caption{Stellar mass ($M_*$, top), cold gas mass ($M_{gas}$, middle) and the gas fraction, $M_{gas}/M_*$ (bottom), averaged for satellite galaxies in radial bins, as a function of the cluster-centric radius divided by the virial radius of the cluster. The black dashed line shows the average position of the accretion shock in these clusters, while the blue, green and red solid lines show the results for the \citet{Bower06}, \citet{Font08}, and Shocks models. The blue, green and red dotted lines show the Bower, Font and Shocks models without ram pressure stripping, respectively. The crosses at zero radius indicate the average properties of the central galaxies. }
\label{fig:mass}
\end{figure}

In Figure \ref{fig:mass} we show the resulting stellar mass, cold gas mass and gas fraction profiles of cluster galaxies, excluding the central galaxy, averaged within cluster-centric radial bins. The values of the central galaxies are plotted, for comparison, as points at zero radius. Here we define galaxy clusters as halos whose mass is greater than $10^{14} h^{-1} M_{\odot}$, and we include all galaxies in each radial bin in the averages. It should be noted that the choice of which galaxies to include makes a large difference in the properties that are observed, as can be seen in the significantly different trends obtained in figures \ref{fig:Balogh} and \ref{fig:ver} where we have selected a different set of galaxies to compare to observations.

We compare our Shocks model (red solid lines), with the Font (green solid lines) and Bower (blue solid lines) models. It should be noted that there are other parameter differences between these models: the Shocks model has had a parameter associated with AGN feedback adjusted from its Bower model value to match the local galaxy $b_J$- and $K$-band luminosity functions, while the Font model has a different value of the metal yield to match the zero point colors of the red and blue sequences. Thus, the difference between the Bower model and the Shocks and Font models is due both to their differing treatments of RPS, and to their different physical parameters. To assess the relative contributions of each of these components, we also consider the Shocks, Font and Bower models without RPS, thus isolating the effects of the parameter changes (red, green and blue dotted lines). We also show the position of the average accretion shock radius of these clusters as a black dashed vertical line, and the properties of the central galaxies as crosses. When these values are very different from those of the satellite galaxies, we leave them out of the plot as the satellite galaxy trends are more interesting for the purposes of this work.

In the top panel of Figure \ref{fig:mass}, we see that in all of the models there is a peak in galactic stellar mass in the centers of clusters; as the central galaxy has been excluded, this shows an increase in stellar mass of the innermost satellites in all models. This is reasonable, as we expect more massive satellite galaxies to sink deeper into the potential well of the cluster. Further, other than a slight rise in stellar mass towards the center of clusters in the Font and Shocks models, the average stellar mass of galaxies remains more or less constant with radius, with the Shocks model having the lowest average level of stellar mass, the Bower model slightly more, and the Font model having the most stellar mass of all. The flatness of these curves indicates that most of the stars of satellite galaxies were formed before they merged with the cluster. Note that the central galaxies in all of the models have more than an order of magnitude more stellar mass than the satellite galaxies, as would be expected since they are at the center of the potential well of the cluster and are not subject to RPS.

The distribution of cold gas mass, in the middle panel of Figure \ref{fig:mass}, shows a different trend. The Bower model, with its more extreme removal of gas in clusters, shows the least cold gas in satellite galaxies, with a very distinct drop in gas mass at around $1.5$ times the virial radius. This drop in cold gas mass is quite steep, with its slope determined by the timescale of satellite orbits as compared to the timescale on which cold gas is made into stars in the satellite galaxies. Interestingly, the Shocks model shows a very similar trend with slightly more gas mass inside the radius of the accretion shock, as would be expected. That the Bower model also shows a drop at around this radius, which happens to be the average accretion shock radius, is interesting, and points to the `preprocessing' of satellite galaxies, in which satellite galaxies experience weaker environmental effects as members of smaller groups of galaxies before merging with the cluster. The Bower model is likely to exhibit stronger group effects than the Shocks and Font models, since its RPS efficiency is always very high, while the RPS of the Shocks and Font models depend on the density of the halo intracluster medium (ICM). Thus, we see that these two very different models of environmental effects give a similar qualitative prediction for the radial dependence of satellite gas mass, indicating that if such a trend in gas mass is observed, we cannot distinguish between a model whose RPS starts at the accretion radius and a much stronger RPS model which starts at the virial radius. 

The Font model predicts a higher level of gas mass at all radii, as is reasonable due to its less harsh RPS implementation, while as expected the models without ram pressure exhibit significantly higher gas mass at all radii than the other models, since it is RPS that is mainly responsible for the sharp drop in gas mass with decreasing cluster-centric radius. It is reasonable that there is a slight drop-off in gas mass towards the center of the cluster even in these models, since in these models the accretion of new gas from the intergalactic medium (IGM) onto the hot gaseous halo of the satellites is suppressed though none of their halo gas is removed. We see that the central galaxies have higher gas mass than satellites, as they are not affected by RPS and also gain gas through mergers. The higher level of gas mass in central galaxies in the models without RPS is likely due to merging satellites, which have not had gas removed by RPS and therefore give more gas mass to the central galaxy.

Finally, in the bottom panel of Figure \ref{fig:mass} we can see the average ratio of cold gas to stellar mass in satellite galaxies. It is quite striking that, despite the differences in gas mass and stellar mass when averaged separately, the average of their ratio is nearly the same in all of these models, with the only significant difference occurring between the Bower model and the Font and Shocks models at less than half of the virial radius. The general behavior of this quantity, in all of the models, is a gradual increase in gas fraction moving into the cluster until the virial radius, at which the gas mass declines precipitously. This peak is likely due to the biased sample of galaxies located close to massive clusters. As they are in general more massive, they may well still be cooling gas and thus forming stars at rates higher than further away from the cluster.

Further inwards, the Shocks and Font models reach a higher, roughly flat, central gas fraction than the Bower model, as expected due to their less extreme removal of gas in cluster members. We also see that the gas fraction reaches only a slightly larger value in the runs without any RPS, indicating that at least the subtle ram pressure of the Shocks and Font models makes little difference in the gas fraction of cluster galaxies, while the implementation of a Bower-type complete removal of gas makes a far larger difference. Note, finally, that the spike in the Bower model gas fraction very close to the center is due to a single galaxy in our sample, which formed in the final timestep and so exhibits an unusually low stellar mass and therefore a very high gas fraction. Its location deep within the cluster is surprising given its very recent formation and it may represent a flaw in the halo detection or tree building algorithms. We leave it in the sample in any case for completeness. Finally, the gas fractions of central galaxies in these models are all well below those of satellites, which is reasonable as cluster central galaxies tend to be ellipticals which have little cold gas although they may have large hot gas halos.

\begin{figure}[t!]
\includegraphics[width=8.5cm]{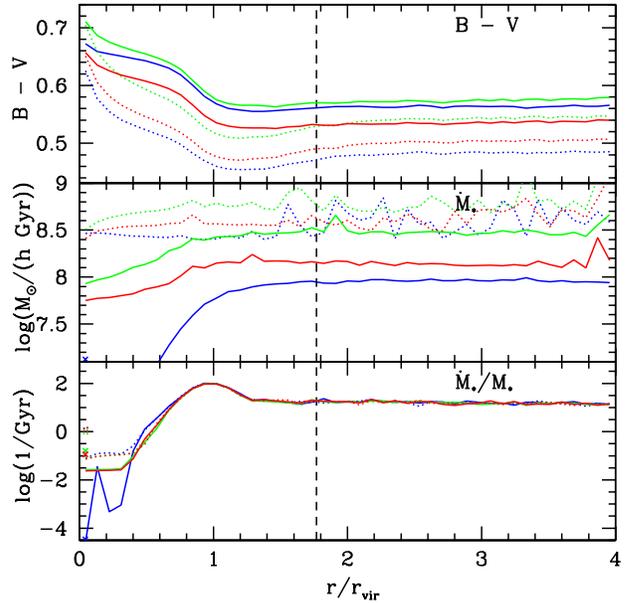}
\caption{B-V colors (top), star formation rate (middle), and specific star formation rate (bottom)), averaged for satellite galaxies in radial bins, as a function of the cluster-centric radius divided by the virial radius of the cluster. The lines and models are the same as in figure \ref{fig:mass}.}
\label{fig:color}
\end{figure} 

In Figure \ref{fig:color}, we show observationally measurable quantities related to the implementation of RPS: the B$-$V color, SFR, and specific SFR of satellite galaxies in clusters. Each of these quantities, as in the previous figure, is calculated for each galaxy and then averaged within cluster-centric radial bins. 

In the top panel of Figure \ref{fig:color} we can see that in all of the models the satellite galaxies become increasingly redder moving towards the center of the cluster as expected, with this reddening beginning around the virial radius and having the same general shape in all models. We see that the Shocks model has bluer colors than the other models, and that the Font model has slightly redder colors than the Bower model. The Font model, with the most gas, has the most dust extinction, and it is this effect which makes it redder than the Bower model. As the separation between the Shocks and Font models is preserved when RPS is turned off, we see that these average color differences result from the star formation and AGN feedback parameter adjustments which were required to bring the models into agreement with the local luminosity function.  As expected, without RPS we see much bluer colors in all of the models. The central galaxies are all much redder than their satellites, which follows from our knowledge that these galaxies tend to be large elliptical galaxies.

In the middle panel of Figure \ref{fig:color}, we see the SFR variation with cluster-centric radius. The SFR versus radius for the Shocks and Font models decreases moderately inside the virial radius, in a similar manner to the gas mass, though not quite as steep. We see a distinct decrease in star formation in the Bower model towards the center of the cluster, due to its harsher implementation of RPS and thus lower gas mass. The Font model has more star formation at all radii with respect to the other models, while the Bower model always has the lowest SFR. This is reasonable as it mirrors the order of strength of RPS in these models. Also as expected,  the models without RPS show higher rates of star formation than the other models. The central galaxies in these models have much larger SFRs than their satellite galaxies, which is reasonable given their larger gas mass.

In the lower panel of Figure \ref{fig:color}, we see the average specific SFR, that is the average of the ratio of the SFR and the total stellar mass of satellite galaxies. The trends in this plot are very similar to those of the gas fraction plotted in the lower panel of figure \ref{fig:mass}. This is no coincidence; in fact the SFR in \galsp is determined such that 

\be \frac{\dot{M}_*}{M_*} = \frac{\epsilon}{\tau_{\rm disk}} (1-R) \frac{M_{gas}}{M_*}, \ee

where $\epsilon$ is the star formation efficiency, $\tau_{\rm disk}$ is the disk timescale, and $R$ is the fraction of mass going into stars that is recycled back into the interstellar medium. As with the gas fraction, a single outlier galaxy in the Bower model is causing the spike at low radius, and can be ignored. We see that the central galaxies in all models have significantly higher specific SFR, as expected. Also, the specific SFR is directly proportional to the gas fraction, and as $\epsilon$ and $R$ are constants and $\tau_{\rm disk}$ does not depend strongly on environment, we see similar behavior in the specific SFR as a function of cluster-centric radius as we saw in the gas fraction. As with the gas fraction, it is very similar in all models and shows a very slightly rising level moving in towards the virial radius, at which there is a peak and then a sharp decline towards the center of the cluster. As with the gas fraction, the models without RPS reach a higher central value of the specific SFR, the Font and Shocks models are very similar, and the Bower model shows a much lower central specific SFR. The central galaxies in these clusters all show a much lower specific SFR than the satellites; this is directly related to their low gas fraction, which is understood as these galaxies tend to be large ellipticals.

\section{Comparison with Observations}  \label{sec:obs}

We compare the results of our \galsp model including accretion shocks and the more extreme RPS model of \citet{Bower06} to recent observations and simulations. In section \ref{sec:Balogh}, we compare with the simulations of \citet{Balogh00} and their observational dataset selected from the CNOC1 cluster redshift survey \citep{Yee96}. In section \ref{sec:Verdugo} we compare the Shocks model to the cluster spectroscopy of \citet{Verdugo08}. In both papers, cosmological parameters $\Lambda = 0.7$, $\Omega_0 = 0.3$, and $h = 0.7$ are used. Despite the fact that {\sc Galform} uses a different set of cosmological parameters as listed in section \ref{sec:Galform}, we conduct our comparison analyses using the same parameters as the observations so as to better reproduce their analysis. In both comparisons, we plot the properties only of satellites in halos with masses greater than $10^{14} h^{-1} M_{\odot}$ to more accurately mimic the selection of massive clusters in these two samples.

\subsection{Comparison to CNOC1 Cluster Redshift Survey as Selected by \citet{Balogh00}}  \label{sec:Balogh}

\begin{figure}[t!]
\includegraphics[width=8.5cm]{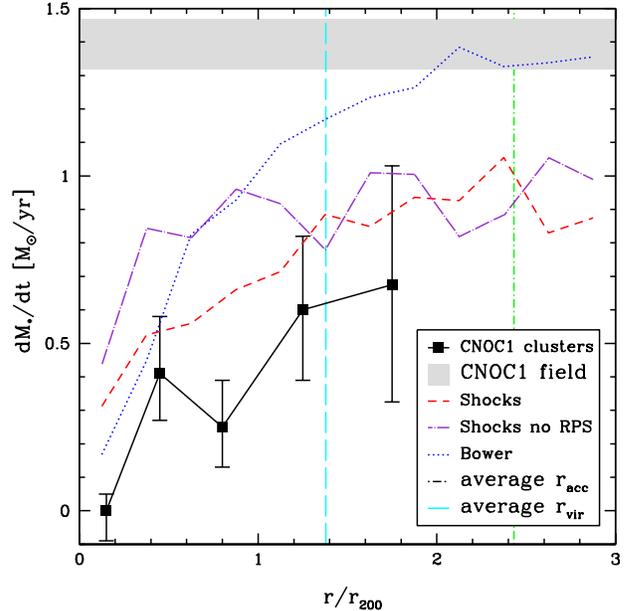}
\caption{Comparison of the cluster redshift survey CNOC1 data as obtained by \citet{Balogh00}, with the Millennium results using three different implementations of RPS. The CNOC1 data is shown as black squares and solid lines, while the extreme ram pressure stripping model of \citet{Bower06} is shown in blue dotted lines, the more nuanced accretion shocks model of this paper is shown in red dashed lines, and this same Shocks model without ram pressure stripping is shown in purple dot-dashed lines. The vertical lines show the average virial radius (cyan long-dashed) and accretion shock radius (green short dot-dashed) for the clusters considered in our semi-analytic model. Note that the x-axis of this plot is the projected cluster-centric radius divided by $r_{200}$ as defined in the text.}
\label{fig:Balogh}
\end{figure}

The cluster redshift survey CNOC1 \citep{Yee96} provides spectra for member galaxies of 15 X-ray luminous clusters. We compare our simulations to the sample of twelve of these clusters presented in \citet{Balogh00}, which were selected to lie within a redshift range $0.19~<~$z$~<~0.45$ and have well-defined cluster centers. Cluster members were selected based on velocity and magnitude cuts. The cluster-centric radii of galaxies were measured relative to $r_{200}$, the radius at which the interior density is $200$ times the critical density. See \citet{Balogh00} for a complete description of the selection criteria.

To ensure accurate comparison to the data, we mimic the observational techniques as described in section 2 of \citet{Balogh00} in our analysis of the Millennium/\galsp models. We analyze clusters at redshift $z=0.3$, and cluster members are selected based on projected radial position, magnitude, and velocity. Specifically, we select galaxies whose projected cluster-centric radius is less than $2~r_{200}$, where $r_{200}$ is calculated assuming an NFW-type density profile, with a Gunn r-band magnitude greater than $-18.8 +5\log h$ at z$=0.3$, and with velocity within $3~\sigma$ of the average velocity of all previously selected galaxies, including the effect of Hubble expansion. The average virial radius (as defined by {\sc Galform} using the overdensity of a spherical top-hat collapse model for this cosmology) of the clusters we ``observe" in this manner is $1.37~r_{200}$.  As in \citet{Balogh00}, we remove the central cluster galaxies from the sample.

We calculate the SFR as was done with the observations, by using the equivalent width of the \OII\ line as a direct indicator of SFR. \citet{Balogh00} find a prescription to determine SFR from observed line widths and luminosities that matches the relationship between these quantities that they see in their semi-analytic simulations. In their prescription, the SFR is the product of this equivalent width with the rest-frame B-band luminosity, an extinction factor, and a normalization constant chosen by comparison with their simulations. In Figure~\ref{fig:Balogh} we show the comparison of the CNOC1 data of \citet{Balogh00} with the results of taking our semi-analytically determined cluster galaxy luminosities and line widths and plugging them into this prescription. We note that this prescription tends to under-predict the SFRs of our galaxies as compared to the rates directly obtained from \gal, and that this effect is strongest for models with less RPS, more galactic gas and dust and thus more extinction. Thus, the SFRs inferred from this method agree fairly well with the directly computed rates for the Bower model, but are much lower than the directly computed rates of the Shocks model. This shows that prescriptions to determine SFR from observed quantities are quite model dependent, and in this case disagree with the relationship we see in \gal.

The resulting curves are compared with the CNOC1 data in Figure \ref{fig:Balogh}. Here we plot three models of ram-pressure stripping: the \citet{Bower06} model, our new Shocks model, and this same Shocks model with the RPS turned off (blue dotted, red dashed and purple long-dot-dashed lines, respectively). We also show, with cyan long-dashed and green short-dot-dashed lines, the average virial and accretion shock radii of the clusters considered. The radius plotted on the x-axis is the projected cluster-centric radius. Note that the SFRs shown in figure \ref{fig:Balogh} are different from those in figure \ref{fig:color} due to their differing definitions of radius, different selection of galaxies, the different units and the linear scale in figure \ref{fig:Balogh} as compared to the logarithmic scale in figure \ref{fig:color}. When taking these into account, the SFRs in the two plots are similar.

We can see that all three models predict similar radial trends in SFR. The Bower model, with more extreme RPS, is quite close to the data in the innermost regions, but has a much steeper rise to larger radii than is observed, thus overproducing the SFR in the outer regions. The Shocks model rises more gradually in the outer regions of the cluster, bringing both of these models into better agreement with the data than the Bower model at large radii. However, we can see that, given the large error bars on the data, neither the Shocks model (with RPS) or the Bower model is a significantly better fit to the observations. This interesting result implies that, even if only RPS is acting to cause these gradients in SFR, from data such as these alone we cannot distinguish the details of the active RPS mechanism. Nevertheless, our Shocks model is marginally the best match to the data.

\subsection{Comparison to \citet{Verdugo08} Cluster Spectroscopy}  \label{sec:Verdugo}

\begin{figure}[t!]
\includegraphics[width=8.5cm]{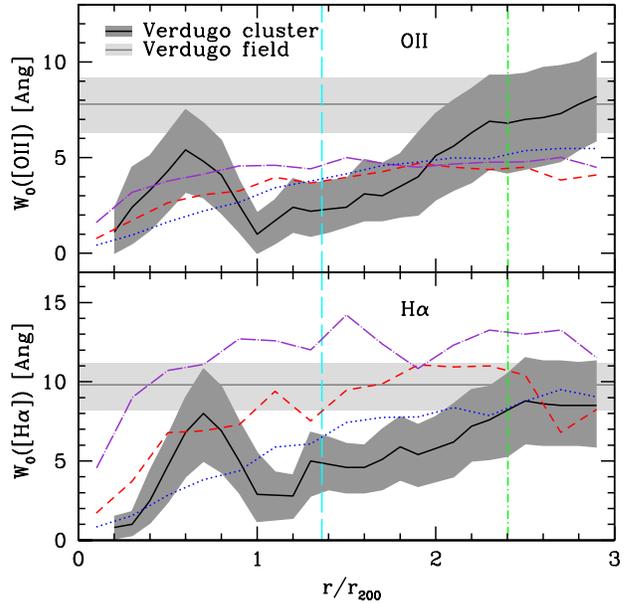}
\caption{Comparison of the \citet{Verdugo08} cluster spectroscopy with our semi-analytic results using three different implementations of ram pressure stripping. The x-axis is the projected cluster-centric radius divided by $r_{200}$, which is defined differently than in the previous section, as explained in the text. The solid black lines with dark grey shading show the observed cluster results and their errors, while the solid grey lines surrounded by light gray show the observed results and errors for field galaxies. The models shown here are the same as in Figure \ref{fig:Balogh}.}
\label{fig:ver}
\end{figure}

We also compare with the sample of six galaxy clusters spectroscopically observed by \citet{Verdugo08}. The star formation activity of cluster galaxies as a function of cluster-centric radius was investigated using the equivalent widths of the \HA\ and \OII\ emission lines, and a significant drop in star formation in cluster galaxies relative to the field was observed. For the details of the observations, the reader is referred to \citet{Verdugo08}.

To mimic the observational selection and analysis, we determine the mean cluster redshift and the cluster velocity dispersion using the bi-weight location and scale estimators of \citet{Beers90}, and select cluster members as galaxies whose redshift is within $3 \sigma$ of the mean cluster redshift. We use the projected cluster radius normalized to $r_{200}$, where $r_{200}$ is here defined to be 

\be r_{200} = \frac{\sqrt3}{10} \frac{\sigma}{H(z)}. \ee

We further remove from our cluster galaxy sample those galaxies whose apparent I-band magnitude is above the survey spectroscopic limit $I = 19.5$, those whose line of sight velocity relative to the cluster center places them clearly outside of the cluster, and those whose absolute I-band magnitude is greater than the limit of $M_I = -21.4$, which was imposed on the observed data to treat clusters at different redshifts with the same luminosity restrictions.  

The comparison of the observational results of \citet{Verdugo08} with our ram pressure implementations is shown in Figure \ref{fig:ver}. The solid black lines with dark grey shading show the observed cluster results and their errors, while the solid grey lines surrounded by light grey show the observed results and errors for field galaxies. To this we compare our models; as in Figure \ref{fig:Balogh}, we show the Bower model (blue dotted), and the Shocks model with (red dashed) and without (purple dot-dashed) RPS. We also show the average virial radius (cyan long-dashed) and accretion radius (green short dot-dashed) for the Millennium clusters considered.

In Figure \ref{fig:ver} we can see that, as in Figure \ref{fig:Balogh}, the average virial radius of these clusters is around $1.4~r_{200}$, while the average accretion shock radius is at $2.5~r_{200}$.  This is just at the edge of the radial extent of the observations, and indicates that if our model of the accretion shock is correct, then the current observations are not probing the cluster beyond the radius of environmental effects. 

The top panel of Figure \ref{fig:ver} shows the radial dependence of the \OII\ equivalent width. We see that the three models predict very similar, nearly straight-line radial profiles, all of which are plausible given the size of the observational error bars. Therefore, these cluster galaxy \OII\ data cannot reliably determine whether any form of RPS is causing the observed decline in SFR. However, all of the models tend to predict a shallower slope than is observed.

In the lower panel of Figure \ref{fig:ver}, we see the radial profile of the \HA line equivalent width. In this case, both the Bower and Shocks models are plausible with the given error bars, while the Bower model predicts a lower overall equivalent width and fits the data slightly better.

It can be seen that the data from both \citet{Verdugo08} and \citet{Balogh00} have a local star formation peak at around $0.5~r_{200}$. The statistical significance of these peaks, due to the size of the errors, is questionable, and their correlation can not be immediately understood to be due to the same phenomenon, as $r_{200}$ was defined differently in these two papers. However, if these peaks do in fact represent a general feature of cluster members' star formation rates, then this effect is the product of physics which has not been included in the current model.

\section{Discussion and Conclusions: Implications for Observations of Ram Pressure Stripping}  \label{sec:disc}

We have addressed the question of the nature of the environmental effects felt by satellite galaxies in galaxy clusters by implementing three different prescriptions of RPS onto the Millennium Simulation, using the semi-analytic galaxy formation model of \gal. The three models implemented are the complete RPS of \citet{Bower06}, in which all of the hot gaseous halo of a galaxy is removed when it is first identified as a member of the friends-of-friends group of a more massive halo, the more nuanced model of \citet{Font08}, in which hot halo gas is only stripped up to a maximum stripping radius determined by the orbit of the satellite, and our new model which incorporates the nuanced Font RPS model but differs from their model by using the radius of the cluster accretion shock as the location at which these effects begin and by utilizing the actual satellite orbit measured directly from the N-body simulation.

Considering the results of applying these models to the detailed merger histories provided by the Millennium Simulation, we see that, in general, all of the models show similar trends in stellar and gas mass with cluster-centric radius, with differences in normalization due to the different ram pressure models and due to the different parameters adopted in the models to obtain a close fit to the properties of the local galaxy population. Further, the gas fraction and therefore the specific SFR is very similar in all of the models considered, with differences only apparent in the regions less than half of a virial radius from the center of the cluster. We see that, in particular, the Shocks and Bower models predict very similar radial average gas mass profiles, with a sharp down-turn at $1.5$ virial radii, the average radius of the accretion shock in these clusters. This indicates that if we were to observe the gas mass trend in cluster galaxies, we would not be able to distinguish these two very different implementations of RPS.

As expected, we find that the B-V colors of satellite galaxies in all of the models become redder towards the center of the cluster, and that in all of the models we considered these effects begin just outside the virial radius of the cluster. The onset of this reddening well outside of the virial radius, as well as the decrease in gas mass beyond than the virial radius, indicate either the preprocessing of satellite galaxies before their accretion onto clusters or the presence of satellites on highly elliptical orbits which have already passed through the central regions of the cluster.

We have compared the results of our accretion shock model of RPS and that of \citet{Bower06} with the observational results of \citet{Verdugo08} and those of \citet{Yee96} as selected by \citet{Balogh00}. We see in these comparisons that both of these models are consistent with the observations, given their uncertainties. Thus, we conclude that current observational data on the radial SFR gradient in clusters do not strongly discriminate between different models of cluster environmental effects. However, our models clearly indicate that the presence of RPS has a strong effect on radial trends within clusters and we expect that it will be possible to constrain the details of these models given future observations with larger samples of cluster galaxies.

Recent work on the semi-analytical modeling of satellite galaxy stripping effects in clusters was carried out by \citet{Weinmann09}. As in this work, they have modified previous models, in which all halo gas was stripped as soon as a galaxy became a satellite in a larger halo \citep{DeLucia07}, to include a more nuanced form of gas stripping. They model cluster-environment effects entirely with tidal stripping, and presume that gas is lost from satellite galaxies in proportion to the loss of dark matter. The baryonic physics of RPS is neglected in the model of \citet{Weinmann09}, while in the model of this work we neglect tidal stripping. Of course, in cosmological structure formation both of these mechanisms have an effect on the stripping of gas from satellite galaxies. However, using the virial scaling relations, we find that the ratio of the forces of ram pressure and tidal stripping is given by

\be \frac{F_{RPS}}{F_{TS}}\bigg|_{r_{vir}} = A \left(\frac{M_{host}}{M_{sat}}\right)^{2/3}, \ee

\noindent where $A$ is a constant of order unity, so the force of RPS larger than that of tidal stripping, for a satellite at the virial radius, by a factor proportional to the ratio of the mass of the host halo to the mass of the satellite to the two-thirds power. By this argument then, at least at the virial radius RPS exerts a larger force on gas in a satellite halo, and so we can be confident that we have included the dominant physical effect in the present work. The results of \citet{Weinmann09} are compatible with our results, and they also find that several of their models are plausible given the error bars of the data with which they compare.

In conclusion radial trends in galaxy properties around clusters can now be accurately predicted by the {\sc Galform} model of galaxy formation. Measurements of these trends therefore have the potential to place strong constraints on the processes of mass accretion and star formation, both of which are key components of our picture of galaxy formation. We have not discussed the distribution of cluster gas in detail in this work, but it is clearly a key ingredient in any model invoking RPS as a driver of cluster galaxy evolution. Recently, \citet{Bower08} described a more advanced calculation of cluster gas physics within {\sc Galform} which aimed to match the X-ray properties of clusters. Future work in this subject should clearly explore both cluster galaxy and X-ray properties in tandem to ensure realistic modeling of the cluster physics. This, coupled with larger samples of cluster galaxies would greatly improve the statistical power of this method as an important constraint on galaxy formation physics. 

\vspace*{2 mm}

\acknowledgements
We thank the anonymous referee for insightful comments. LGB acknowledges the support of the NSF Graduate Fellowship Program. AJB acknowledges the support of the Gordon \& Betty Moore Foundation. The authors would like to thank the \galsp team, Carlton Baugh, Richard Bower, Shaun Cole, Carlos Frenk and Cedric Lacey, for allowing us to use \galsp in this work. Additionally, we thank Andreea Font for implementing the original calculation of ram pressure stripping in this model.

%
%
%
\bibliography{ref}
%
%
%
%
%
\appendix
\section{Appendix A: Accretion Shock Calculation}

The accretion shock model of \citet{Voit03} upon which we base our implementation makes several assumptions. The cluster is assumed to be spherically symmetric and in hydrodynamic equilibrium, with an effective equation of state  $P(r) \propto [\rho(r)]^{\gamma_{\rm eff}}$, where $\gamma_{\rm eff} = 1.2$, and with the actual equation of state of a free monatomic gas, $P = K(s) \rho^{5/3}$, where K is a function of the specific entropy. The cluster potential is assumed to be of the NFW \citep{Navarro97} type, and the post-shock velocity of the gas is assumed to be negligible.

With these assumptions, they obtain the following hydrostatic model of the gas within the accretion shock: 

\be T(x) = T_{\Delta} g(x) \label{eqn:T} \ee

\be \rho(x) = \rho_g [g(x)]^{1/(\gamma_{\rm eff} - 1)} \label{eqn:rho} \ee

\be P(x) = \frac{T_{\Delta} \rho_g}{\mu m_p} [g(x)]^{\gamma_{\rm eff}/(\gamma_{\rm eff} - 1)} \ee

\be g(x) = g_0(x) + g_1 \label{eqn:g} \ee

\be g_0(x) = \frac{2(\gamma_{\rm eff} - 1)}{\gamma_{\rm eff}} \frac{\ln(1 + c x)}{\ln(1 + c) - c(1 + c)^{-1}} \frac{1}{x}. \label{eqn:g0} \ee

Here, $x \equiv r/r_{vir}$ is the ratio of the cluster-centric radius to the virial radius, $\rho_g$ and $g_1$ are constants of integration, which can be specified by the constraint that the baryonic mass inside the shock radius is equal to the total baryonic mass of the halo $f_b~M$ and the shock jump condition relating post-shock temperature to the incoming velocity, respectively.

In our implementation of this model we relax two particularly unjustified assumptions, that the accretion shock is always strong, and that the constant $g_1$ in equation \ref{eqn:g} is negligible. To relax these assumptions we apply the further (better founded) assumption that the gas accreting onto the halo is compressed and heated adiabatically as it makes its way to the shock radius from the mean density IGM. Using this assumption, and also assuming that the temperature of the mean IGM is 3000 K, we can calculate the temperature just outside the shock, from which we can obtain the Mach number of the shock $\mathcal{M}$.

With this prescription for finding the strength of the shock $\mathcal{M}$, where $\mathcal{M} \to \infty$ is the strong-shock limit initially imposed by \citet{Voit03}, we use the shock relations of \citet{LL} to obtain an equation for $g_1$: 

\be g_1 = \frac{((2\gamma\mathcal{M}^2 - (\gamma - 1))((\gamma - 1)\mathcal{M}^2 + 2)} {\gamma (\mathcal{M}^2 - 1)^2} \frac{\xi}{x_{\rm ac}} - \frac{2(\gamma_{\rm eff} - 1)}{\gamma_{\rm eff}} \frac{\ln(1 + c x_{\rm ac})}{ln(1 + c) - c(1 + c)^{-1}} \frac{1}{x_{\rm ac}}. \label{eqn:g1} \ee

Here, the ratio of the accretion shock radius to the virial radius is shown as $x_{ac}$. We finally find that the equation for the accretion shock, equation A12 in \citet{Voit03}, is modified to

\be \left[\frac{3 \mathcal{C}}{4} \tilde{\rho} I(x_{\rm ac}, g_1, \gamma_{\rm eff}, c)\right]^2 x_{\rm ac}^3 + x_{\rm ac} - 2 = 0, \ee

where $\mathcal{C}$ is the shock compression factor, the ratio of the densities internal and external to the accretion shock, 

\be \mathcal{C} \equiv  \frac{\rho_2}{\rho_1} = \frac{(\gamma + 1)\mathcal{M}^2}{(\gamma - 1)\mathcal{M}^2 + 2},\label{eqn:rhojump} \ee

$I$ is the integral

\be I(x_{\rm ac}, g_1, \gamma_{\rm eff}, c) \equiv \frac{1}{x_{\rm ac}^3} \int_0^{x_{\rm ac}} \left[\frac{g_0(x)+g_1}{g(x_{\rm ac})}\right]^{1/(\gamma - 1)} x^2 dx, \ee

and  we use the variable $\tilde{\rho}$ which is defined in \citet{Voit03} to be

\be \tilde{\rho} \equiv \frac{4}{3} \frac{2}{\Delta}^{1/2} (H t)^{-1} \frac{d \ln M}{d \ln t}. \label{eqn:rhotilde} \ee

There is a further subtlety in the calculation of the mass accretion rate onto halos, which has a strong effect on the final accretion shock radius through its appearance in equation \ref{eqn:rhotilde}. We have explored several possibilities, but in this work we adopt a spherically symmetrized model averaged over the dynamical time of the halo. More specifically, we calculate the average mass accretion rate over the past one dynamical time of the halo, including all of the mass added to the most massive progenitor of the current halo through mergers or accretion. Since we use this mass as an estimate of the spherically accreted mass of the halo, we are making the assumption that all of this added mass, even that which was added through mergers, will distribute itself and cause accretion shocks much as the same mass of spherically symmetric accretion would.

\end{document}